\documentclass[a4paper]{PoS}

\title{The VLA Low Band Ionospheric and Transient Experiment (VLITE): A Commensal Sky Survey}

\ShortTitle{VLITE}

\author{\speaker{Tracy Clarke}, Namir Kassim, Emil Polisensky, Wendy Peters\\
        Remote Sensing Division, Naval Research Laboratory, Washington, DC USA\\
        E-mail: \email{tracy.clarke@nrl.navy.mil}, \email{namir.kassim@nrl.navy.mil}, \email{emil.polisensky@nrl.navy.mil}, \email{wendy.peters@nrl.navy.mil}}

\author{Simona Giacintucci\\
     Computational Physics Inc., Springfield, VA USA\\
       E-mail: \email{simona.giacintucci.ctr@nrl.navy.mil}}

\author{Scott D.\ Hyman\\
     Sweet Briar College, Dept. of Engineering and Physics, Sweet Briar, VA USA\\
       E-mail: \email{shyman@sbc.edu}}

\abstract{The US Naval Research Laboratory (NRL) and the National Radio Astronomy Observatory (NRAO) have collaborated to develop, install, and commission a new commensal system on the Karl G. Jansky Very Large Array (VLA). The VLA Low Band Ionospheric and Transient Experiment (VLITE) makes use of dedicated samplers and fibers to tap the signal from 10 VLA low band receivers and correlate those through a real-time
DiFX correlator. VLITE allows for the simultaneous use of the VLA to observe primary science using the higher frequencies receivers (1$-$50 GHz) through the NRAO WIDAR correlator and lower frequencies through the DiFX correlator. VLITE operates during nearly all observing programs and provides 64 MHz of bandwidth centered at 352 MHz. The operation of VLITE requires no additional resources from the VLA system running the primary science and produces an {\it ad-hoc} sky survey. 

The commensal system greatly expands the capabilities of the VLA through value-added PI science, stand-alone astrophysics, the opening of a new window on transient searches, and serendipity. In the first year of operation we have recorded more than 6300 hours spread across the sky. We present an overview of the VLITE program, discuss the sky coverage and depth obtained during the first year of operation, and briefly outline the astrophysics and transients programs. 

}

\FullConference{EXTRA-RADSUR2015 (*)\\
		20--23 October 2015\\
		Bologna, Italy

                \bigskip
                \hrule
                \bigskip

                \textnormal{(*) This conference has been organized
                  with the support of the Ministry of Foreign Affairs
                  and International Cooperation, Directorate General
                  for the Country Promotion (Bilateral Grant Agreement
                  ZA14GR02 - Mapping the Universe on the Pathway to
                  SKA)}
}

\begin{document}

\section{Introduction}

The US National Radio Astronomy Observatory's Karl G.\ Jansky Very Large Array (NRAO VLA) is a 27 element interferometer located on the Plains of San Augustin in New Mexico. The 25 m dish elements are arranged across three arms oriented in a 'Y-shaped' pattern. The dishes are movable on rail tracks to provide four primary configurations with maximum baselines of 1, 3.4, 11, and 36 km for the D, C, B, and A configurations respectively. The VLA has eight Cassegrain receivers for the primary observing frequencies which provide continuous frequency coverage from 1 to 50 GHz. These feeds are located in a feed ring near the center of the main dish and are selected by motion of the subreflector. A more detailed description of the VLA and the recent upgrade to broadband Cassegrain feeds is provided by \cite{perley11}. 

Investigation of frequencies below 1 GHz on the VLA began in the 1980's in the 330 MHz band and was expanded to include the 74 MHz band in the 1990's. Both systems were in regular use as part of the VLA until the broadband upgrade of the VLA. The low frequency systems were relatively narrow band with only the 330 MHz band offered for regular use while the 74 MHz band was only available for certain campaigns due to the loss of sensitivity at higher frequencies when the dipoles were mounted and in position. For a more detailed description of the development of the low band systems see \cite{kassim07} and references therein.

The US Naval Research Laboratory (NRL) collaborated with NRAO to replace the legacy low band receivers with a new broadband four channel receiver that can cover 50$-$500 MHz. The two populated channels expand the legacy low band system and cover the 58$-$84 MHz and 230$-$470 MHz regions. This system has been in regular use since 2013 (see \cite{clarke11} and \cite{clarke16} for more details).

The location of the low band receivers near the prime focus of the VLA antennas offers a unique opportunity to develop a commensal low band system that works in parallel with the higher-frequency Cassegrain feeds. The primary goals for the VLA Low Band Ionospheric and Transient Experiment (VLITE\footnote{See \href{http://vlite.nrao.edu}{vlite.nrao.edu} for more details on VLITE.}) are to demonstrate: 1) the power of a sensitive, continual ionospheric monitor that provides fluctuation spectra of ionospheric variations for analysis of climatological trends and rare events, 2) the capabilities of a fully commensal low frequency system on a cm-wave instrument to expand the phase-space for radio transient phenomena, and 3) the scientific payoff across a broad range of low frequency astrophysics using a cm-wave instrument in a piggy-back mode. Here we concentrate on the final two goals (for ionospheric applications see \cite{joe}); we describe this new system and discuss initial work on astrophysics and transients. Finally, we discuss how VLITE can be used as a pathfinder for a future 27-antenna, broad-band "LOw Band Observatory" (LOBO) commensal system on the VLA. 

\section{VLITE}

VLITE provides real-time correlation of the signal from a subset of the low band receivers on the NRAO VLA. On antennas equipped with the VLITE samplers, the signal from the 230$-$470 MHz channel of the low band receiver is split to allow the signal to pass to either the NRAO WIDAR correlator or the VLITE correlator. Along the VLITE path, the X and Y polarization signals pass through the VLITE module in the antenna vertex room. This module has 8-bit samplers clocked at 1024 MHz. The output is formatted in the VDIF format (\cite{vdif}) and transported on spare NRAO fiber to the VLITE correlator located in the VLA control building. The VLITE data are centered at 352 MHz with a total bandwidth of 64 MHz although we note that the usable frequency range is limited to 320$-$360 MHz due to strong radio frequency interference (RFI) in the upper portion of the band.

The signals from the antennas are correlated in a custom DiFX (\cite{deller07}) correlator that was modified to include real-time network-based VDIF correlation, 8-bit VDIF data, and linear polarization. Currently VLITE correlates data from 10 antennas, producing four correlation products with 100 kHz wide channels between 320 and 384 MHz and 2 second integration time. VLITE records and correlates all VLA scans of length roughly 6s up to one hour. Scans shorter than 6s total are currently lost due to initial setup needed for each scan while scans longer than one hour (which are very rare) are truncated at one hour. The original VLITE deployment is scheduled to be a minimum of a two year operation through at least two VLA A and B configurations. VLITE operations have minimal impact on NRAO resources and no impact on resources available to the principle investigator programs running on the telescope. 

\section{VLITE Sky Coverage}

\begin{figure}[t]
	\centering{
     \includegraphics[width=.7\textwidth]{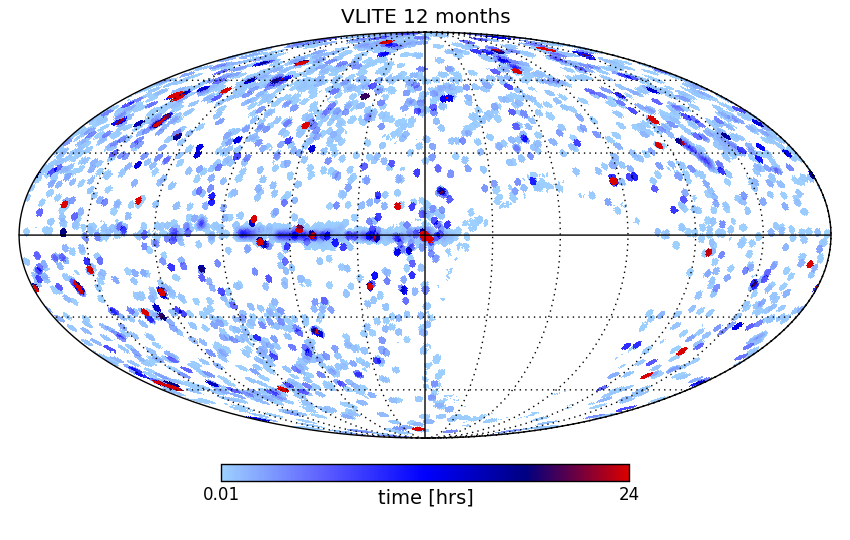}}
     \caption{Sky coverage for first year of VLITE operation plotted on an Aitoff projection in Galactic coordinates. The red regions show deep accumulated fields such as COSMOS, the Galactic center, and calibrators.}
     \label{figure:fig1}
     \end{figure}

Science operations for VLITE began on 25 November 2014 during the VLA C configuration. In the first year of operation VLITE recorded a total of 6300 hours of VLA time, equivalent to 71\% wall-clock time. Observations are spread across the sky for calibrator and target fields. We show in Figure~\ref{figure:fig1} the first year sky coverage for VLITE plotted on an Aitoff projection in Galactic coordinates. The majority of the small red regions are calibrators which are routinely used. We note that although these calibrators may be observed using different primary observing bands, we calibrate each band separately in VLITE and the final data products can be combined across all primary setups to produce much deeper images. Other fields with total integration times exceeding 24 hours include the Galactic center, the COSMOS field (\cite{Smolcic}), and the Hubble Ultra Deep Field (\cite{beckwith}). The empty region in the lower right represents the southern sky which is not visible to the VLA.

\begin{figure}[t]
	\centering{
     \includegraphics[width=.6\textwidth]{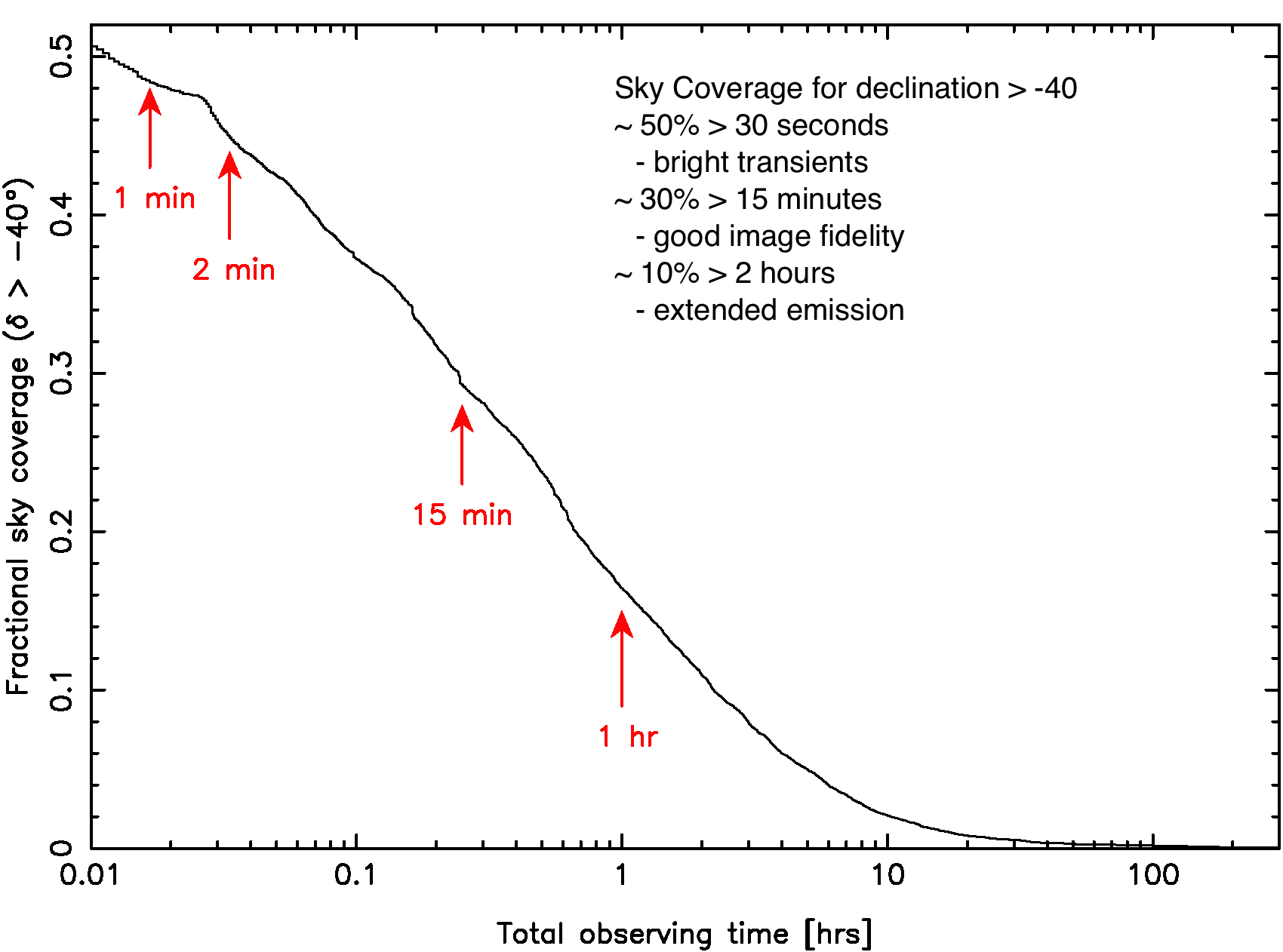}}
     \caption{Plot of the fractional sky coverage above declination of -40$^\circ$ as a function of total observing time obtained from data acquired during the first year of VLITE operation.}
     \label{figure:fig2}
     \end{figure}

In Figure~\ref{figure:fig2} we show the distribution of the observing time during VLITE's first year of operation as a function of the fractional sky coverage above declinations of -40$^\circ$ for scans longer than 36 seconds. We have data for roughly 50\% of this region of the sky at the shallowest depth. These images allow us to build a catalog of the bright sources in these regions of the sky. Measured source fluxes can be used for long-term monitoring of bright, compact sources as well as a sky model and quiescent reference grid to use in the search for very bright transients. For longer integrations of order 15 minutes, the first year of VLITE provides coverage over roughly 30\% of the sky. The longer integrations provide much better image fidelity with fewer imaging artifacts and increased sensitivity to extended structures. Very deep integrations, exceeding 2 hours, are available for more than 10\% of the sky at $\delta$ $>$ -40$^\circ$. These images provide the best image fidelity and are sensitive to faint compact sources as well as extended diffuse emission.

Image quality depends on factors beyond simply the integration time. The primary factor impacting image quality outside the Galactic plane is the filling factor of the $uv$ plane. Short total integration time that is spread over a large hour angle coverage has much better image fidelity than the same amount of time obtained consecutively. Within the Galactic plane there are calibration difficulties due to the significant differences between the 1.4 GHz NVSS (\cite{condon}) survey used as a sky model during VLITE calibration and the true sky emission within the lower frequency VLITE band. The rms level depends also on the VLA configuration. We note that for the majority of the initial VLA A configuration VLITE was in a more compact configuration with the outermost antennas located just beyond the standard B configuration outer pad along each arm. This B+ configuration for VLITE provided the best rms while still allowing for good image quality. Typical rms levels for the VLITE B+ configurations for integration times of 6 minutes, 1 hour, and 10 hours are 10 mJy/beam, 5 mJy/beam, and 1 mJy/beam respectively, while for the compact D configuration the rms levels are a factor of $20-30$ higher due to confusion.

\section{Astrophysics with VLITE}

VLITE data are processed for astrophysical programs through an automated calibration and imaging pipeline that runs one day following data acquisition. This pipeline is based on the Obit data reduction software (\cite{obit}) and runs through standard steps of delay calibration, RFI excision, bandpass and flux calibration, imaging and (if feasible) up to two rounds of phase self-calibration. The pipeline data products include calibrated $uv$ data, calibration tables, image cubes, spectral index and spectral curvature maps, broadband total intensity images, and a source catalog. The initial pipeline processes only produces Stokes I images although additional polarization processing may be feasible once a better understanding is developed for the polarization properties of the VLA low frequency band.

\begin{figure}[t]
     \includegraphics[width=.6\textwidth]{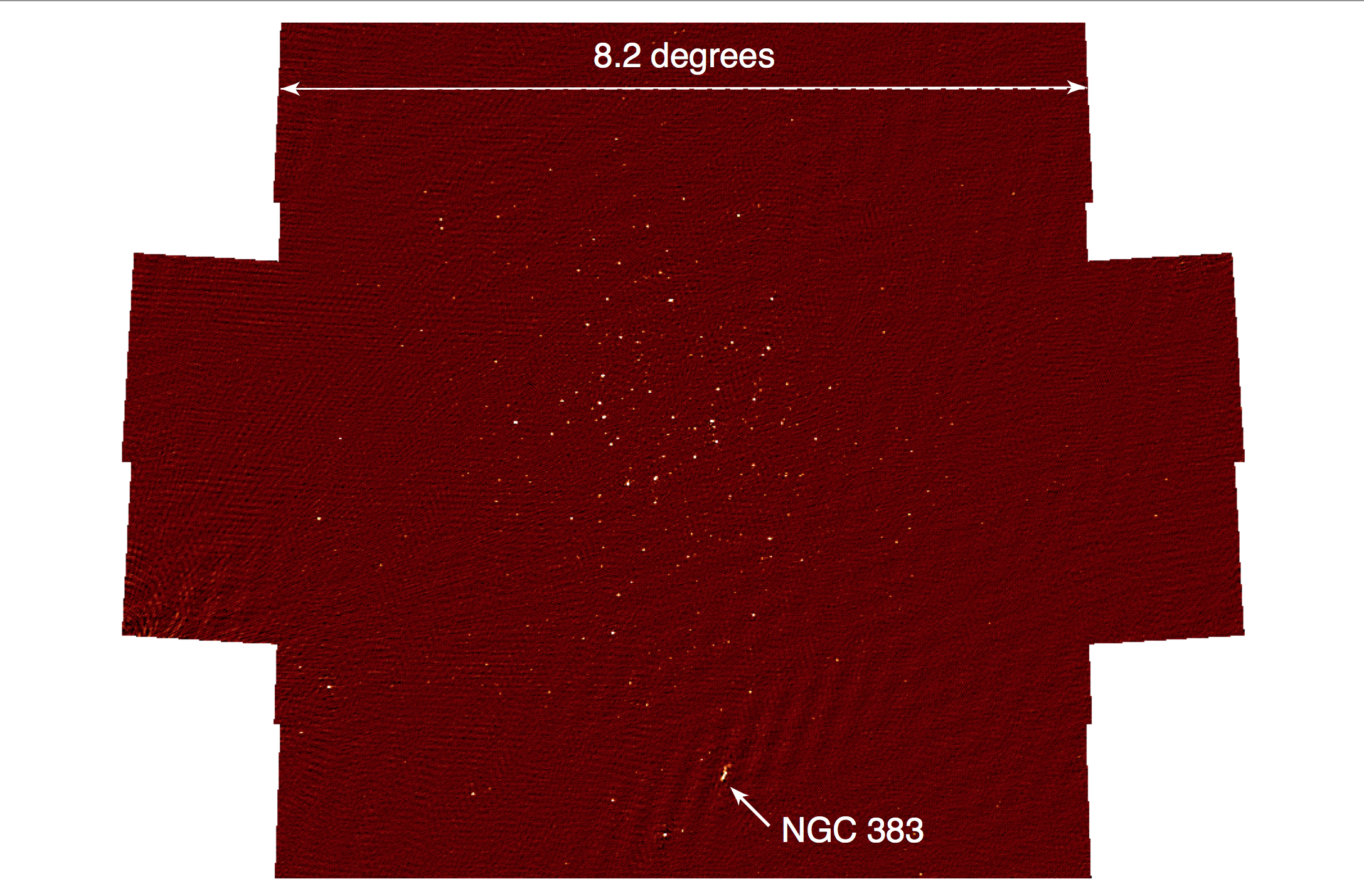}
     \includegraphics[width=.38\textwidth]{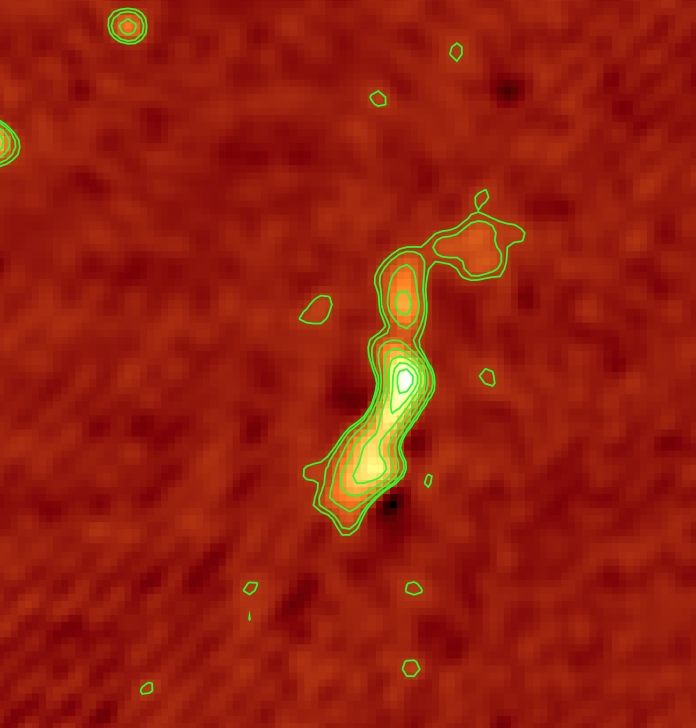}
     \caption{Left panel shows the VLITE image of a field observed 01 December 2014 showing a field of view roughly 8 degrees across. Bright regions show the sources within this field. The giant radio galaxy NGC383 (right panel) is detected roughly 3.5 degrees from the phase center.}
     \label{figure:fig3}
     \end{figure}

In the left panel of Figure~\ref{figure:fig3} we show an example of an 8$^\circ$ diameter VLITE field obtained during the C configuration while in the right panel we show details of the source NGC 383 located nearly 3.5 degrees from the phase center. The image was processed manually for comparison with the final astrophysical pipeline image. The very large field of view is possible because of the broad low frequency beam created by the out-of-focus location of the feeds on the VLA. These large images can be used for study of source morphology or to search for bright transient events but we caution that accurate flux measurements will likely only be feasible within 2$^\circ$ of the phase center where the central lobe of the beam is well behaved. 

The calibration and image quality from the VLITE astrophysics pipeline has been compared to the manual processing of data from all configurations to both refine the pipeline as well as understand where end-users may improve on pipeline data products. Overall the VLITE pipeline calibration is very robust and matches very well to manual calibration. The majority of science programs will be able to directly use the final pipeline calibration products. The imaging and self-calibration undertaken by the pipeline have several underlying assumptions that are necessary to process large volumes of commensal data. In some cases, we have found that minor additional RFI excision followed by self-calibration can improve the rms on VLITE images by factors of up to a few. We note that for the majority of well-behaved fields outside the Galactic plane, there is little gain from additional manual processing. Within the Galactic plane, VLITE image quality can be significantly improved through use of more accurate sky models.

The astrophysical applications for VLITE data include value-added science for the primary VLA program  where VLITE data can be used for morphological, spectral, and/or variability work to enhance the scientific results. VLITE data can also be used for stand-alone science where the large fields of view allow for detection and characterization of large samples of sources. We also note that VLITE data can be used to form an inhomogeneous sky catalog of sources above declinations of -45$^\circ$.

\section{Commensal Transient Hunting}

Astrophysical transients at low radio frequencies often fall in the regime of coherent emission from energetic events with very high brightness temperature (e.g.\ Solar bursts, electron cyclotron maser, pulsars). Plots of the phase-space occupied by the coherent transients (\cite{cordes}) show that the known systems cover a wide range of durations and brightness although they are generally limited to fairly short durations. Incoherent transient emission (e.g.\ flare stars, supernovae, X-ray binaries), on the other hand, tend toward much longer timescales at low radio frequencies (\cite{Fendera}).  Due to the large sky coverage and long timescales, VLITE is well suited to searches for transients in a very wide range of phase space across both the coherent and incoherent transient regimes.

One approach to transient detection with VLITE involves comparison of sources within the VLITE catalog to currently available deep catalogs such as the NVSS (\cite{condon}) and WENSS (\cite{Rengelink}). As VLITE operates longer, it will be possible to build much deeper sky models from VLITE data by combining overlapping data in areas of the sky. These can be used both as an improved quiescent sky model for identifying new transients, as well as for identification and characterization of slowly-varying emission which would be characteristic of distant incoherent transient sources.

The VLITE transient processing pipeline utilizes similar calibration and imaging strategies as the astrophysics pipeline. Our initial testing and development of the transient search program uses the LOFAR TraP software (\cite{swinbank}). During the recent outburst of the black-hole binary system V404 Cyg, we were able to use the TraP to identify this source as a transient with high confidence by VLITE (\cite{kassimv404}). 

\begin{figure}
\centering{
     \includegraphics[width=.7\textwidth]{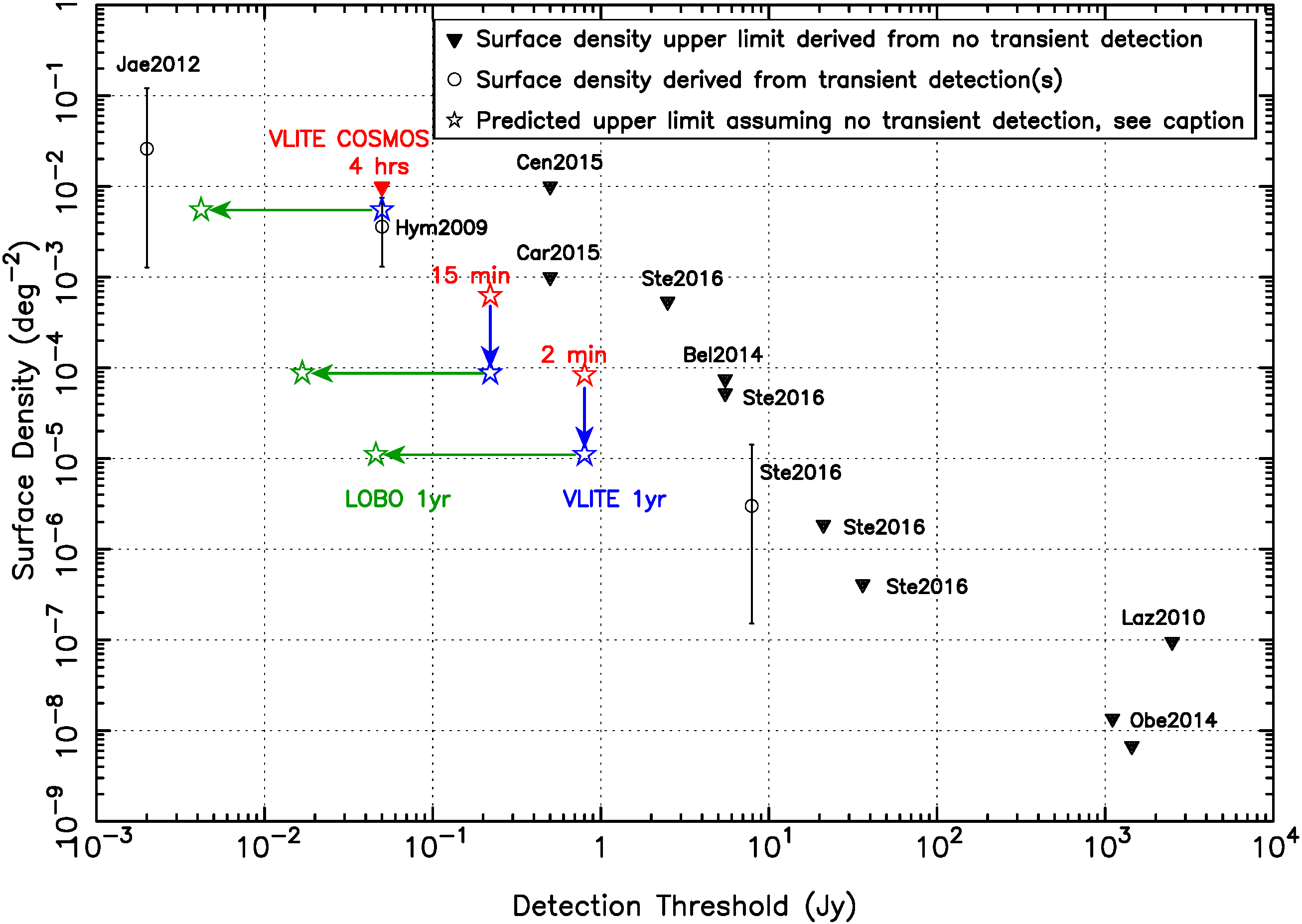}}
     \caption{Radio transient surface densities against the survey detection threshold for blind searches of transients at sub-GHz frequencies. Black triangles indicate upper limits from no transient detections. Round points with 95\% confidence bars are the surface density bounds from surveys with transient detections. Red points show the VLITE upper limits from a monitoring campaign of the COSMOS field.  Star points indicate predicted upper limits when the images are divided into shorter exposures. Blue points are the expected limits from no detections in all exposures in one year of VLITE data. In green are the one year predictions for a full 27 antenna, full bandwidth LOw Band Observatory. (Figure adapted after \cite{Fendera} and \cite{Stewart}.)}
     \label{figure:fig4}
     \end{figure}
Efforts at many low frequency instruments are currently directed at blind surveys for {\it radio-selected} transients by taking advantage of the naturally large fields of view. VLITE's early foray into this regime leveraged two deep VLA campaigns approximately centered on the COSMOS (\cite{Smolcic}) field. The TraP software was used to search a succession of 55 $\times$ 4 hr B configuration images (1$\sigma$ $\sim$ 5 mJy/beam in each image) obtained approximately nightly during February-May 2015 (55 days of observations spread over 74 calendar days). Figure~\ref{figure:fig4} (filled red triangle) shows the VLITE surface density limit of 9.9 x $10^{-3}$ deg$^{-2}$ at 50 mJy (10$\sigma$), derived from the analysis of the 4 hr images. The next step is to sub-divide these 4 hr images into 15 and 2 minute sub-images, whose limits (assuming no detection) are indicated by the red stars in Figure~\ref{figure:fig4}. Projecting forward to our complete analysis of the first year of data, the blue stars in Figure~\ref{figure:fig4} indicate the improved limits that we anticipate for ALL 4 hr, 15 min, and 2 min exposures. We note that this would include comparison across VLITE images and with existing catalogs.

\section{Toward the LOw Band Observatory: LOBO}

VLITE development was undertaken with a vision of transitioning the narrow-band, 10 antenna system into a new VLA capability across all 27 antennas and covering the full low frequency bandwidth. This new LOw Band Observatory (LOBO) would run similar to VLITE on all observing programs but would be a much more powerful system, providing significantly increased imaging fidelity, much lower rms noise levels, and broad spectral coverage.  These characteristics of LOBO would make it a powerful astrophysics tool for morphological and spectral study of faint low frequency source populations and extended diffuse structure. To demonstrate the transient capabilities of LOBO, we show in Figure~\ref{figure:fig4} the expected transient surface density limits for 1 year as green stars. From its combination of both sky coverage and sensitivity, LOBO clearly contributes to the concerted international efforts to explore the sub-GHz sky for radio transients.

\section{Acknowledgments}

Basic research in radio astronomy at the Naval Research Laboratory is supported by 6.1 Base Funding. This work made use of data from the VLA Low-band Ionospheric and Transient Experiment (VLITE). Construction and installation of VLITE was supported by the Naval Research Laboratory (NRL) Sustainment Restoration and Maintenance funding. The National Radio Astronomy Observatory is a facility of the National Science Foundation operated under cooperative agreement by Associated Universities, Inc.

\end{document}